\documentclass[preprint,aps,prb]{revtex4}

\usepackage{epsf}
\usepackage{bm}

\begin{document}
    
\title{Finite temperature behavior of impurity doped 
                     Lithium cluster {\em viz} Li$_6$Sn. }

\author{Kavita Joshi and D. G. Kanhere}

\affiliation{%
Department of Physics, University of Pune, 
Ganeshkhind, Pune 411 007, India}

\date{\today}

\begin{abstract} 

        We have carried out extensive isokinetic {\it ab initio} molecular
dynamic simulations to investigate the finite temperature properties of
the impurity doped cluster Li$_6$Sn along with the host cluster Li$_7$.  
The data obtained from about 20 temperatures and total simulation time of
at least 3~ns is used to extract thermodynamical quantities like canonical
specific heat. We observe a substantial charge transfer from all Li atoms
to Sn which inturn weakens the Li-Li bonds in Li$_6$Sn compared to the
bonds in Li$_7$. This weakening of bonds changes the finite temperature
behavior of Li$_6$Sn significantly.  Firstly, Li$_6$Sn becomes liquid-like
around 250~K, a much lower temperature than that of Li$_7$
($\approx$~425~K).  Secondly, an additional quasirotational motion of
lithium atoms appears at lower temperatures giving rise to a shoulder
around 50~K in the specific heat curve of Li$_6$Sn. The peak in the
specific heat of Li$_7$ is very broad and the specific heat does not show
any premelting features.

\end{abstract}

\pacs{31.15.Ar,36.40.Ei,36.40.Sx}

\maketitle

%
%
\section{Introduction\label{sec:intro}}

       Small clusters are known to behave differently than their bulk
counterparts, mainly because of their finite size. The finite size effects
are reflected in the most of their properties, like equilibrium
geometries, energetics, optical properties, ionization potential,
polarizabilities, etc\cite{ref1,ref2,ref3,ref4,ref5}.  For example,
equilibrium geometries of small as well as medium size clusters are
considerably different than the bulk structures. Stability of a class of
clusters depends on the filling up of geometric or electronic shell. The
well known example of such a stability is that of alkali metal clusters
displaying magic numbers. Further, the properties of homogeneous clusters
may be altered, some times quite significantly by adding an impurity.  In
the last few years some work has been reported on the impurity doped
clusters\cite{heteroclust,heteroclust1}.  These studies have brought out a
number of interesting aspects like traping of an impurity, modifications
in the equilibrium structures, changes in the bonding characteristics and
enhancement or otherwise in the stability. A few studies have also been
reported on the optical properties. Many of these properties have been
shown to get influnced by the relative difference in the valence, ionic
radii and electronegativity.

	A majority of the investigations on heterogeneous and homogeneous
clusters pertain to the ground state properties.  During the last few
years a number of experimental\cite{Naexpt,tinexpt} and
theoretical\cite{melttheo,sn10,sn20} studies on finite temperature
behavior of clusters have been reported. Finite temperature behavior of
homogeneous clusters turns out to be very intriguing and has brought out
many interesting features in contrast to the bulk behavior\cite{amvrev}.  
For instance, clusters exhibit a broad peak in the specific heat which is
sometimes accompanied by a premelting feature like a shoulder indicating
possible isomarization.  Recent experiments\cite{tinexpt} and
simulations\cite{sn10,sn20} on tin clusters have shown that these clusters
become liquidlike at higher temperatures than the bulk melting temperature
(Tm$_{{\rm bulk}}$) contrary to the belief that solidlike to liquidlike
transition in the finite size systems occurs at a temperature lower than
Tm$_{{\rm bulk}}$ because of the surface effects. A number of experiments
on Na clusters by Haberland and co-workers\cite{Naexpt} have demonstrated
that there is no systematic size dependence observed in the melting
temperature of clusters between sizes 50 to 200. More interestingly, the
peaks in the melting temperature as function of size do not occur at the
geometric or electronic shell closing.

        In the present work, we investigate the finite temperature
behavior of Li$_6$Sn and Li$_7$. Since the impurity is known to change the
geometry as well as bonding substantially in the host cluster it is also
expected to change the finite temperature properties of the host cluster.
The motivation for investigating these clusters comes from our previous
study of Sn doped Li clusters\cite{linsn}. This study has shown that,
doping of lithium cluster with a tetravalent impurity such as Sn, induces
a charge transfer from Li to Sn and as a result the nature of bonding
changes from metallic like to dominantly ionic like. Preliminary
investigation of the finite temperature behavior of Li$_6$Sn has also
indicated that this charge transfer induces dramatic effects at low
temperatures. The interest in these systems also stems from the anomalous
behavior seen in the bulk properties ({\em e.g.} electrical resistivity,
density, stability function) of Li-Group IV alloys as a function of
composition\cite{Lugt}.  These alloys show peak in the resistivity around
20\% Sn composition and this has been interpreted as atomic clustering
around Sn at this composition whereas other alkali metal-Group IV alloys
show peak in the resistivity around 50\% composition of Group IV elements.
It has also been observed that Li-Si alloys\cite{lisi} show 15\% volume
contraction in the liquid state for Si concentration between 15\% to 50\%
whereas in Na-Sn alloys\cite{nasn} melting takes place via {\em rotor}
phase in which alkali atoms are suppose to diffuse freely while the group
IV polyanions perform only rotational or vibrational movement.

As we shall demonstrate in this paper, the impurity induces a
quasirotational motion of alkali atoms around Sn at low temperatures. In
addition, in the liquid-like state the value of root mean square bond
length fluctuations ($\delta_{{\rm rms}}$) of impurity doped cluster turns
out to be considerably lower than that of Li$_7$ indicating that volume
expansion upon melting is much less in this cluster. Although it is
difficult to establish a direct correlation between the properties of the
bulk alloys and a small cluster like Li$_6$Sn, we believe that the physics
of impurity induced effects in both the systems essentially has the same
origin arising out of the charge transfer which results into the strong
ionic bond.

It is now well established that for reliable ion dynamics one must
reproduce the bonding properly.  Description of ion dynamics critically
depends on the correct microscopic description of instantaneous state of
electrons as system evolves in time. This is especially true when the
nature of bonding may change during evolution of the trajectories at
different temperatures. Therefore, for obtaining reliable ion dynamics we
have chosen to use density functional theory(DFT). Small size of the
system has permitted us to perform {\it ab initio} calculations for
substantially larger simulation time as compared to the previously
published {\em ab initio} results\cite{melttheo,sn10,sn20}.

The paper is organized as follows.  In Sec.~\ref{sec:comput} we describe
briefly our computational and statistical approaches.  Results for the
equilibrium geometries, nature of bonding and finite temperature
properties are given in Sec.~\ref{sec:results}, and the conclusions are
given in Sec.~\ref{sec:conclusions}.

%
%

\section{Computational Details\label{sec:comput}}   

All the calculations have been performed using the Kohn-Sham formulation
of density-functional molecular dynamics (MD),\cite{Payne} using the
nonlocal norm-conserving pseudo-potentials of Bachelet {\em et al.}\
\cite{BHS} in Kleinman-Bylander form \cite{KB} with the Ceperley-Alder
\cite{CA} exchange-correlation functional.  A cubic supercell of length
30~a.u. is used with energy cutoff of 15~Ry. for Li$_6$Sn and 12~Ry. for
Li$_7$, which is found to provide sufficient convergence in total
electronic energy(E$_{tot}$).  To have a sufficiently precise evaluation
of the Hellmann-Feynman forces on the ions, we ensure that the residual
norm of each KS orbital, defined as $|H_{{\rm KS}}\psi _{i}-\epsilon
_{i}\psi _{i}|^{2}$ ($\epsilon _{i}$ being the eigenvalue corresponding to
eigenstate $\psi _{i}$ of the KS Hamiltonian $H_{{\rm KS}}$), was
maintained at $10^{-12}$~a.u.  After every ionic movement the total
electronic energy is converged up to $10^{-5}$~a.u. for low temperatures
(T$\leq$ 200~K) and the convergence is $10^{-4}$~a.u. for higher
temperatures. Such a high convergence is not warranted in the simulated
annealing runs but it is very crucial for probing finite temperature
behavior of impurity doped clusters. The ground state and other
equilibrium structures have been found by carrying out several
steepest-descent runs starting from structures chosen periodically from a
high-temperature simulation.

The ionic phase space is sampled by isokinetic MD,\cite{isokin} in which
the kinetic energy is held constant using velocity scaling.  For Li$_{7}$
we split the temperature range 100~K $\leq T\leq$ 1000~K into about 20
different temperatures, performing up to 90~ps of simulation for lower
temperatures ({\em i.e.} T $\leq$ 200) and 150-200~ps for higher
temperatures.  For Li$_{6}$Sn we split the temperature range 10~K $\leq
T\leq$ 900~K into about 23 different temperatures, performing up to 200~ps
of simulation for lower temperatures ({\em i.e.} T $\leq$ 200) and 150~ps
for higher temperatures.  This results into total simulation time about
3~ns for Li$_7$ and 5~ns for Li$_6$Sn.

The data obtained from these extensive simulations is used to extract
different thermodynamic quantities such as entropy, specific heat, etc.
The ionic specific heat is calculated from the fluctuations in the
internal energy of the system and is given by, \begin{equation}
C(T)=\frac{3Nk_{B}}{2}+\frac{1}{k_{B}T^2} ( \langle V^2 \rangle_{t} -
{\langle V \rangle}_{t}^2 ) , \label{eqn:Cv_by_f} \end{equation} where $V$
is total electronic energy and $\langle \cdots \rangle _{t}$ denotes time
averaged over the entire trajectory. A smooth specific heat can be
obtained by using multiple histogram (MH) technique\cite{MH}.  With multi
histogram fit, which is basically a least square fit we extract density of
states ($\Omega (E)$) or entropy ($S(E)=k_{B}\ln \Omega (E)$).  From
entropy we can compute various thermodynamic quantities like canonical
specific heat, $C(T)=\partial U(T)/\partial T$, where $U(T)=\int
E\,p(E,T)\,dE$ is the average total energy, and where the probability of
observing an energy $E$ at a temperature $T$ is given by the Gibbs
distribution $p(E,T)=\Omega (E)\exp (-E/k_{B}T)/Z(T)$, with $Z(T) $ the
normalizing canonical partition function.

Along with these thermodynamic quantities we also analyze the data using
tradition parameters such as the $\delta _{{\rm rms}}$, defined as
\begin{equation}
\delta _{{\rm rms}}=\frac{2}{N(N-1)}\sum_{I>J}\frac{(\langle
R_{IJ}^{2}\rangle _{t}-\langle R_{IJ}\rangle _{t}^{2})^{1/2}}{\langle
R_{IJ}\rangle _{t}},  \label{eqn:deltarms}
\end{equation}
where $N$ is the number of ions in the system, $R_{IJ}$ is the distance
between ions $I$ and $J$, and $\langle \cdots \rangle _{t}$ denotes a time
average over the entire trajectory.

\begin{figure}
\epsfxsize=10.0cm
\centerline{\epsfbox{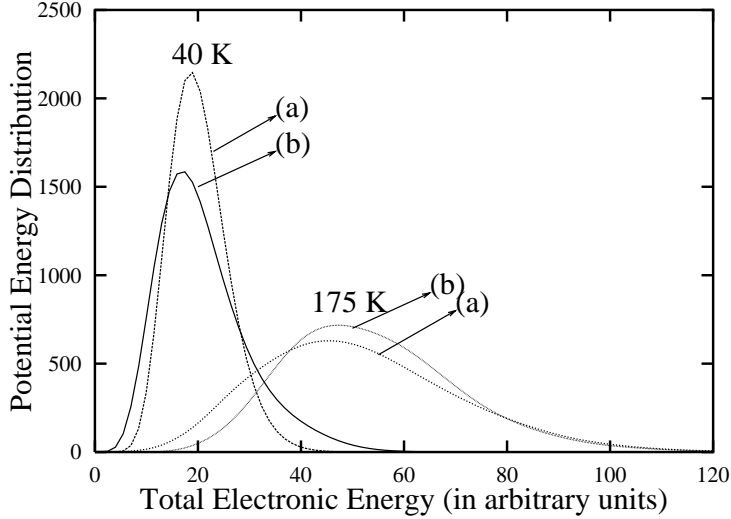}}
\caption{\label{fig0} Comparison of PED for 40~K and 175~K using different 
criteria for convergence in the total electronic energy 
($\Delta E_{{\rm SCF}}$).
{\rm (a)} denotes $\Delta E_{{\rm SCF}}=0.00001$~a.u. and 
{\rm (b)} denotes $\Delta E_{{\rm SCF}}=0.0001$~a.u.}
\end{figure}               

We would like to make few pertinent comments on some technical issues
which are specific to {\it ab initio} simulations. Since such DFT
simulations are quite expensive, a judicious choice of simulation time
step ($\Delta t$) , criteria for the convergence in the total electronic
energy during Born-Oppenheimer MD ($\Delta E_{{\rm SCF}}$) and total
simulation time has to be made.  Some of these issues may not be very
crucial for geometry optimization, since there the details of ion dynamics
are secondary. However for reliable extraction of thermodynamic data, the
ion dynamics should be faithfully simulated using the accurate forces.
Since the calculation of forces involves the total electronic energy
(E$_{tot}$), the choice of $\Delta E_{{\rm SCF}}$ can turn out to be
crucial. We demonstrate this point by showing potential energy
distribution (PED) (shown in Fig.\ \ref{fig0}) calculated using two
different $\Delta E_{{\rm SCF}}$ criteria, 0.0001 and 0.00001~a.u., for
two temperatures 40~K and 175~K. Let us recall that PED(E,T) denotes
accessible energy states between $E$ and $E+dE$ (where E is total
electronic energy of the system). Fig.\ \ref{fig0} clearly shows that two
different $\Delta E_{{\rm SCF}}$ leads to considerably different PEDs. The
height of the peak in curve (a) ({\em i.e.} $\Delta E_{{\rm
SCF}}=0.00001$~a.u.)  is about 25\% higher than the curve (b) ({\em i.e.}
$\Delta E_{{\rm SCF}}=0.0001$~a.u.)  at 40~K. Further, it can be seen that
the lower accuracy curve is seen to span a broader energy range and we
have verified that this leads to the erroneous specific heat in this range
of temperatures. However as the temperature rises lower criteria can be
used without lose of accuracy.

%
%
\section{Results and Discussions\label{sec:results}}

\subsection{Finite temperature behavior of host cluster: Li$_7$
\label{sec:Li$_7$}}

\begin{figure}
\epsfxsize=10.0cm
\centerline{\epsfbox{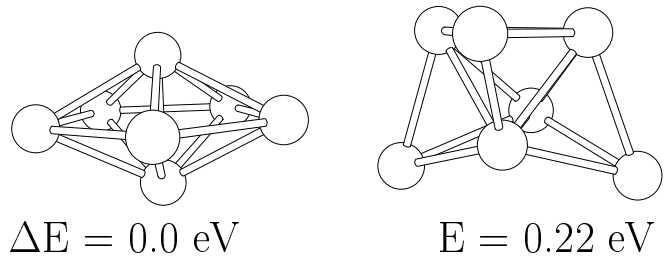}}
\caption{\label{fig1}The ground state and the first excited state 
geometry of Li$_7$. $\Delta$E represents difference in the total energies.}
\end{figure}

We begin our discussion by examining the equilibrium geometries of
Li$_{7}$ (shown in Fig.\ \ref{fig1}). The ground state structure is a
pentagonal bipyramid with D$_{5h}$ symmetry.  The excited state is
tricapped tetragon (C$_{2v}$ symmetry) with $0.22$ eV higher in energy
compared to the ground state.  These geometries are consistent with the
earlier reported work\cite{Li7geo}.  As expected the charge distribution
is completely delocalized indicating metallic like bonding in this cluster
(not shown).

\begin{figure} 
\epsfxsize=10.0cm 
\centerline{\epsfbox{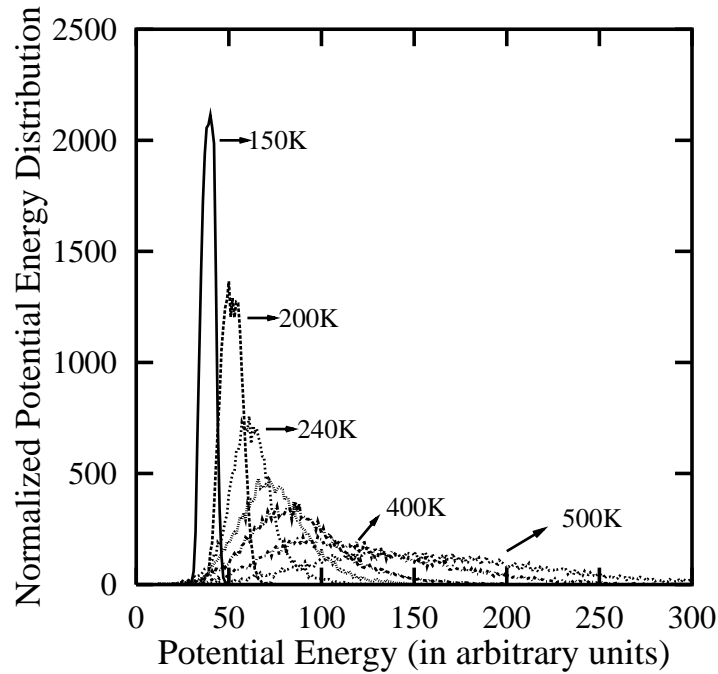}}
\caption{\label{fig2} Normalized potential energy distribution of Li$_{7}$
for some relevant temperatures; {\em viz} 150~K, 200~K, 240~K, 280~K,
320~K, 400~K and 500~K. It should be noted that PED for 400~K and 500~K
are very flat and almost overlapping} 
\end{figure}

Before discussing the thermodynamic quantities like specific heat it is
instructive to examine normalized potential energy distribution (PED)
obtained via our molecular dynamic simulations, as a function of
temperature.  In Fig.\ \ref{fig2} we show PED for different temperatures
ranging from 150~K to 500~K.  It may be noted from Fig.\ \ref{fig2} that
the PED for low temperatures is very sharp and narrow indicating that the
system spans a very restricted range of energies. Indeed the observed
motion at these temperatures shows that atoms oscillate around their
equilibrium positions. As temperature increases the pumped in energy is
used to break some of the bonds and as a result cluster spans wider
configuration space which reflects in the broader PED.  We note that
around 400~K the PED becomes very flat which indicates that the cluster is
spanning large number of energy states with equal probability.  This
rather broad PED around 400~K and above as compared to the PED for lower
temperatures is suggestive of liquid-like nature of the cluster in this
range of temperatures.

\begin{figure} 
\epsfxsize=10.0cm 
\centerline{\epsfbox{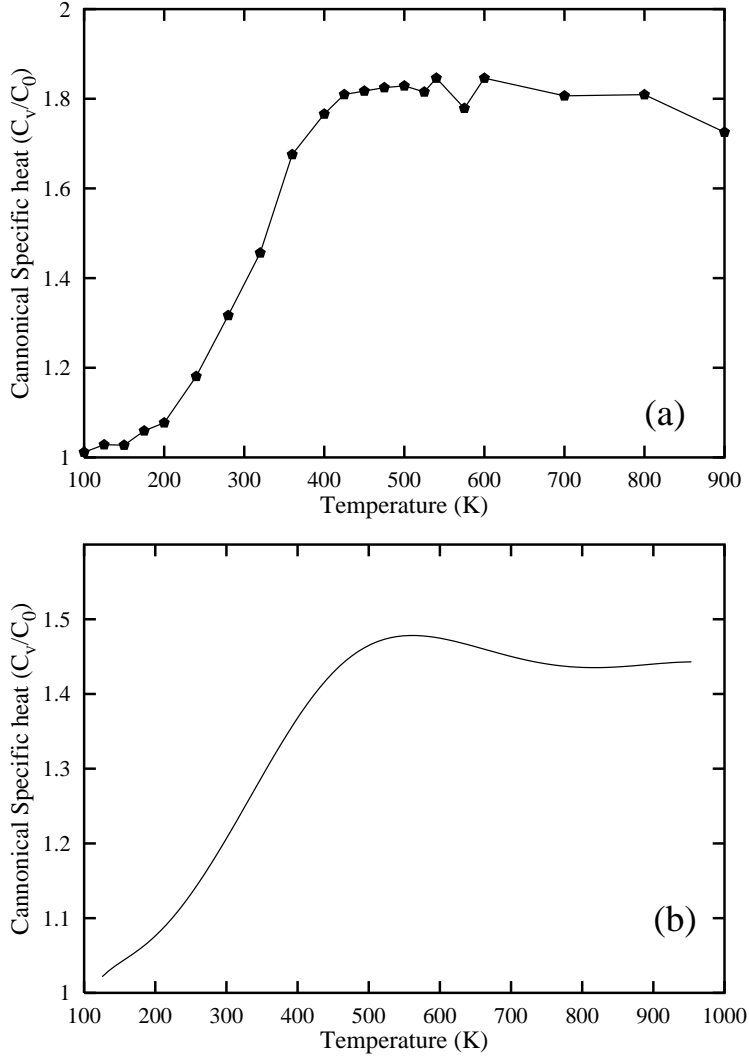}}
\caption{\label{fig3} The canonical ionic specific heat of Li$_{7}$.
({\rm a}): Computed directly from fluctuations using Eq.\ \ref{eqn:Cv_by_f} 
with 75~ps for temperatures up to 200~K and 150~ps or more for higher
temperatures.  ({\rm b}): The specific heat curve calculated using
multihisto fit.} 
\end{figure}

Some insight can be gained by observing the detailed motion (the movie) of
the cluster at few interesting temperature.  We note that, around 360~K
cluster starts revisiting the ground state via the first excited state and
during this process, atoms from the pentagonal ring and atoms capping the
ring get interchanged.  This processes of interchanging of atoms, becomes
more frequent with increasing temperature.  Indeed as we shall see this
behavior leads to the peak in the specific heat around 425~K. In Fig.\
\ref{fig3} we show ionic specific heat of Li$_7$ calculated using Eq.\
\ref{eqn:Cv_by_f} directly (shown in Fig.\ \ref{fig3} a) as well as with
the multiple histogram technique (shown in Fig.\ \ref{fig3} b). Evidently,
both the methods yield similar specific heat except that specific heat by
MH is much smoother. We note that the specific heat is very broad and does
not change significantly after the peak around 425~K. Such a broad peak in
the specific heat is a characteristic of small clusters in contrast to the
sharp transition observed in the extended systems.

\begin{figure}
\epsfxsize=10.0cm
\centerline{\epsfbox{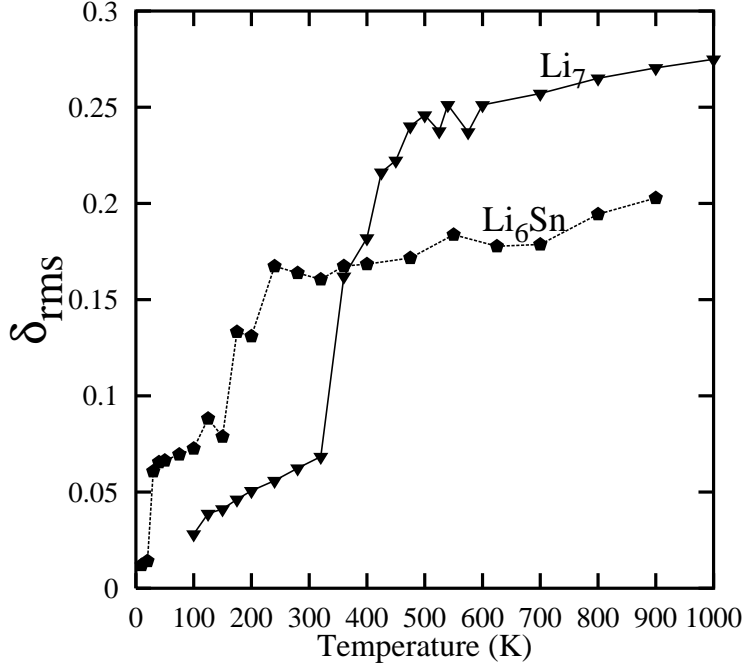}}
\caption{\label{fig4} Root mean square bond length fluctuations of
Li$_{7}$ and Li$_6$Sn. The slow monotonous rise in $\delta_{{\rm rms}}$ 
for Li$_7$ below 320~K indicates solid-like behavior whereas 
linear nature above 450~K shows liquid-like behavior.
For Li$_{6}$Sn the Lindmann criteria gets satisfied at much lower 
temperature (around 175~K).
We also note that for Li$6$Sn the value of 
$\delta_{{\rm rms}}$ remains considerably constant for a wide 
range of temperature {\em i.e.} from 250~K to 700~K.}
\end{figure}               

	These observations are also supported by the $\delta_{{\rm rms}}$
which is shown in Fig.\ \ref{fig4}.  As the name suggest $\delta_{{\rm
rms}}$ indicates fluctuations in the bond lengths averaged over all pairs
and simulation time.  According to the Lindmann criteria the clusters are
considered to be in the liquid-like state when $\delta_{{\rm rms}}$
exceeds 0.1.  It is interesting to note that $\delta_{{\rm rms}}$ shows
three distinct regions, a solid-like region below 320~K and a liquid-like
region above 450~K. 320~K to 450~K is a transition region and the value of
$\delta_{{\rm rms}}$ rises sharply in this region.  If we chose to use the
Lindmann criteria to estimate the ``melting temperature" then this cluster
``melts" between 320~K to 360~K.

	To summarize, the specific heat curve indicates that above 425~K
Li$_{7}$ is definitely {\em liquid-like} and the transition is
characterized by rather broad peak in the specific heat.  These features
are consistent with the specific heat of Li$_8$ reported in Ref.\
[\onlinecite {Li8thermo}]. Both these clusters namely Li$_7$ and Li$_8$ do
not show any premelting features.

\subsection{Finite temperature behavior of impurity doped cluster:
Li$_6$Sn \label{sec:Li$_6$Sn}}

	In this section we present our results on the impurity doped
cluster, namely Li$_6$Sn and contrast the behavior with that of host
cluster Li$_7$.

\begin{figure}
\epsfxsize=10.0cm
\centerline{\epsfbox{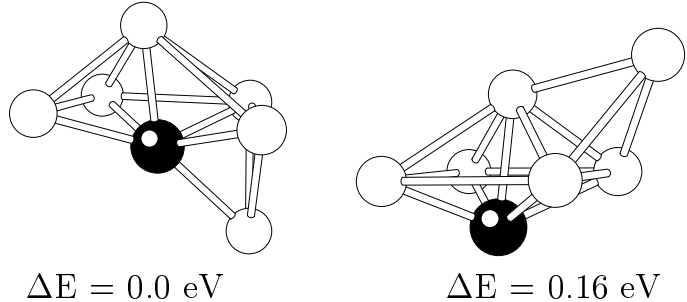}}
\caption{\label{fig5} The ground state and the first excited state 
geometry of Li$_6$Sn. $\Delta$E represents difference in the total energies.}
\end{figure}

	Before studying these finite temperature effects let us examine
the equilibrium structures of Li$_6$Sn which are shown in Fig.\
\ref{fig5}. Firstly, we note that both the geometries are considerably
different than that of Li$_7$. Secondly, in the ground state, Sn has six
fold coordination whereas the first excited state which is 0.16~eV above
the ground state, has five fold coordination with Li atoms. Thus the first
excited state has one less Li-Sn bond. Further, these two structures can
be obtained by capping the most stable Li$_4$Sn tetrahedron \cite{linsn}
in two different ways.

\begin{figure}
\epsfxsize=10.0cm
\centerline{\epsfbox{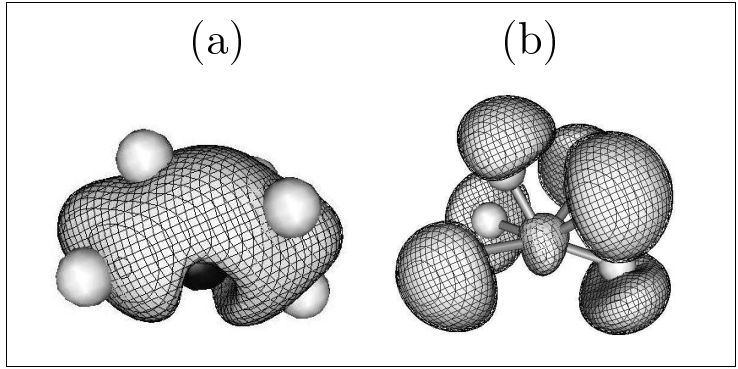}}
\caption{\label{fig6}The isosurfaces for difference charge-density of Li$_6$Sn.
(a): charge gained region, (b): charge depletion region. Sn is shown by
the black sphere at the center whereas white spheres indicate Li atoms.}
\end{figure}               

Now we turn to the bonding features of the system. Our earlier
work\cite{linsn} has shown that a tetravalent impurity like Sn in lithium
clusters changes the delocalized metallic like bonding to dominantly ionic
like.  In Fig.\ \ref{fig6} we show the difference charge density of
Li$_6$Sn, defined as the difference between the selfconsistent charge
density and the superimposed atomic charge density.  Quite clearly, the
charge gained region is mainly around Sn whereas the charge depletion
region is near all Li atoms.  This clearly shows that there is a charge
transfer from all Li atoms to Sn, filling in the Sn centered {\it p}
orbitals\cite{obser}.  Typically the depleted charge is of the order of
$2/3$ electron per Li atom.  As a consequence of the significant charge
transfer from Li to Sn, considerable depletion of charge in the region of
Li-Li bond is observed which results into weakening of Li-Li bond. This
weakening of Li-Li bonds show dramatic effects on the finite temperature
behavior of the cluster.

\begin{figure}
\epsfxsize=10.0cm
\centerline{\epsfbox{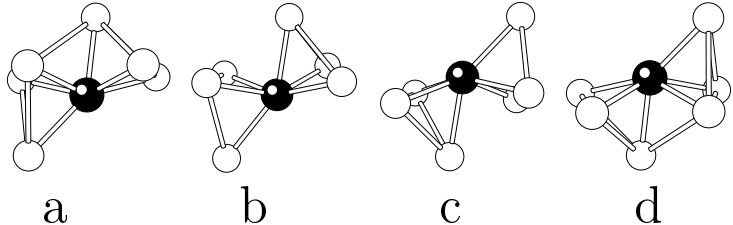}}
\caption{\label{fig7}Quasirotational motion of Li atoms around Sn;
{\rm (a)}: the ground state geometry
{\rm (b and c)}: two Li-Li bonds are broken and the Li atoms get separated 
into two bunches which form a quasirotational motion around Sn.
{\rm (d)}: The resultant geometry. 
We note that {\rm (a)} and {\rm (d)} denotes same geometries with 
different orientation.}
\end{figure}               

\begin{figure}
\epsfxsize=10.0cm
\centerline{\epsfbox{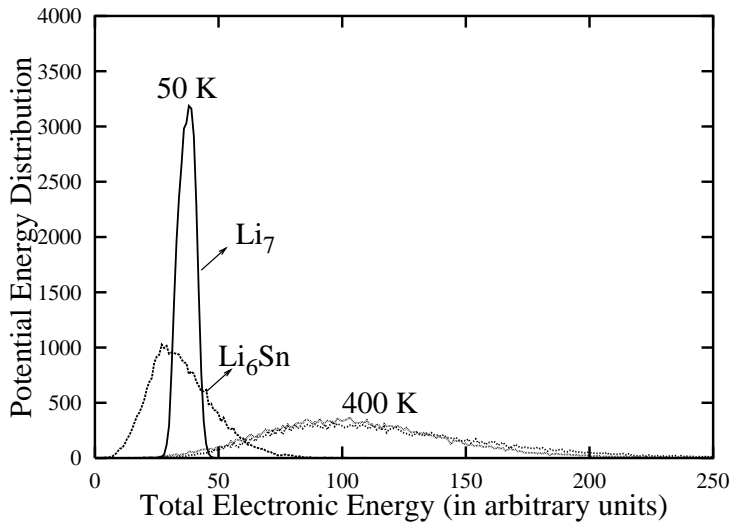}}
\caption{\label{fig8}Normalized Potential Energy Distribution at 
150~K and 400~K of Li$_6$Sn and Li$_7$.}
\end{figure}               

	The consequences of weakening of Li-Li bonds are observed at very
low temperatures. Around 30~K two Li-Li bonds break and Li atoms get
separated into two bunches (see Fig.\ \ref{fig7} b and c ).  These bunches
perform a quasirotational motion around Sn as shown in Fig.\ \ref{fig7}.
We note from Fig.\ \ref{fig7} that the initial (Fig.\ \ref{fig7} a) and
final (Fig.\ \ref{fig7} d) geometries are same but have different
orientations. Since the motion is observed at very low temperatures the
potential barrier joining these two (identical) structures must be very
low. This quasirotational motion seen at such a low temperature shows up
in the PED of Li$_6$Sn. In Fig.\ \ref{fig8} we compare PED of Li$_6$Sn
with that of Li$_7$. Evidently, at 150~K the PED of Li$_6$Sn is
significantly different than that of Li$_7$. A rather narrow and sharp
distribution for Li$_7$ indicates that it is showing clear solid-like
behavior exhibiting small oscillations at this temperature, whereas the
significantly broader distribution of Li$_6$Sn shows that other modes of
excitations are also available to the system at this temperature. In fact
the observed motion at this temperature shows breaking of most of the
Li-Li bonds. On the other hand at 400~K the PED of both clusters
have similar shapes.  

\begin{figure}
\epsfxsize=10.0cm
\centerline{\epsfbox{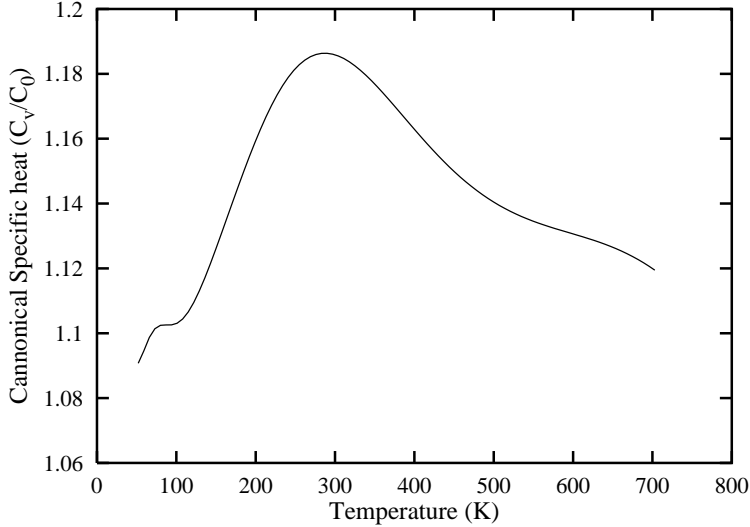}}
\caption{\label{fig9} The canonical ionic specific heat of Li$_{6}$Sn. 
The shoulder around 50~K is because of the quasirotational motion 
seen at that temperature whereas the peak is around 250~K, 
a substantially low temperature compared to that of Li$_7$.}
\end{figure}               

Turning to the ionic specific heat of Li$_6$Sn (shown in Fig.\ \ref{fig9})
we note that the specific heat curve is significantly different than that
of Li$_7$. Unlike Li$_7$, it shows a shoulder around 50~K\cite{footnote}
and the main peak is observed at substantially lower temperature ({\em
i.e.} 250~K ) compared to that of Li$_7$ (425~K). The shoulder in the
specific heat curve of Li$_6$Sn corresponds to the quasirotational motion
described earlier. As temperature rises further this quasirotational
motion becomes more rigorous and eventually around 200~K the bunches of Li
atoms lose their identity and exhibit diffusive motion around Sn.  This
leads to the peak around 250~K in the specific heat.

	These observations are also consistent with the $\delta_{{\rm
rms}}$ which is shown in Fig.\ \ref{fig4} along with the $\delta_{{\rm
rms}}$ for Li$_7$. For Li$_6$Sn the value of $\delta_{{\rm rms}}$ rises in
two distinct steps. The first step is around 30~K. We recall that
quasirotational motion starts around this temperature and another step is
around 200~K where the bunches of Li atoms lose their identity and diffuse
around Sn.

It is also observed that the value of $\delta_{{\rm rms}}$ remains nearly
constant between 250 to 700~K. Although this effect is small it is
noteworthy. In this temperature range the system is {\em liquid-like} but
instead of random diffusive motion of all atoms in the cluster we observe
that Li atoms perform constrained diffusive motion with Sn at the center.  
The averaged fluctuations in Li-Sn bond in this range of temperatures is
found to be within 15\%. Thse observations indicate that the main peak
around 250~K is associated with the breaking of weak Li-Li bonds. We also
note that in Li$_6$Sn though $\delta_{{\rm rms}}$ satisfies the Lindmann
criteria at very low temperature compared to that of Li$_7$, its saturated
value is $0.17$ which is considerably lower compared to the host Li$_7$
($\delta_{{\rm rms}}=0.25$) cluster.  This indicates that in the 
liquidstate the volume expansion of the impurity system is much less
than in the case of host cluster. This is consistent with the existence of
stronger Li-Sn bond in the system.  At still higher temperatures one of
the ionic bonds breaks and we observe the first excited state (shown in
Fig.\ \ref{fig5}) in which Sn has one less coordination number than the
ground state. Thus the route for evaporation proceeds via the first excited
state.

Finally we note that, Li$_6$Sn is an over lithiated cluster, {\em i.e.}
the number of Li atoms in the cluster are more than those required to
complete the {\it p} shell of Sn atom. It will be interesting to
investigate the finite temperature behavior of a cluster which satisfy the
octate rule {\em i.e.} like Li$_{4x}$Sn$_x$. In these clusters though we
expect that because of {\em complete} charge transfer from all Li atoms to
Sn, the bonds between Li atoms will be extremely weak, the ionic bond
between Li-Sn will be more strong than those in the overlithiated cluster
like Li$_6$Sn.

\section {Conclusions\label{sec:conclusions}}

We have presented an {\em ab initio} isokinetic MD simulations on Li$_6$Sn
and Li$_7$ to demonstrate the dramatic effects induced by the impurity in
the host cluster.  The key point which emerges out of these calculations
is that the essential physics governing these effects comes from the
charge transfer from Li to Sn because of the large difference in their
electronegativities. In Li$_6$Sn, we observe a quasirotational motion at
very low temperature which gives rise to a shoulder in the specific heat
curve around 50~K. The main peak is observed at much lower temperature,
around 250~K, compared to the host cluster ($\approx$ 425~K)  whereas
Li$_7$ shows a broad peak in the specific heat with no premelting
features. The motion in the liquid state of Li$_6$Sn is essentially a
constrained motion of Li atoms around Sn center.

\acknowledgments

We gratefully acknowledge the financial support of the ISRO-DRDO, India.


\begin{thebibliography}{}
\bibitem{obser}     We have examined the relevant occupied orbitals and have
                    observed that all the orbitals are centered around Sn.
\bibitem{footnote}  We have verified the exisistance of this small shoulder by 
                    carrying out a number of additional simulations between 
	            10~K to 100~K.
\end{thebibliography}
\end{document}